\documentstyle[sprocl]{article}

\def\Journal#1#2#3#4{{#1} {\bf #2}, #3 (#4)}

\def\be{\begin{equation}}
\def\ee{\end{equation}}
\def\bea{\begin{eqnarray}}
\def\eea{\end{eqnarray}}
\def\complete positivitybar{\hbox{{\rm complete positivity}\hskip-1.80em{/}}}

\begin{document}

\title{TIME SCALE AND  COMPLETELY POSITIVE DYNAMICAL
EVOLUTIONS}

\author{B. VACCHINI}

\address{Dipartimento di Fisica
dell'Universit\`a di Milano and INFN, Sezione di Milano,
\\Via Celoria 16, I-20133, Milano, Italy
\\and
\\Fachbereich Physik, Philipps-Universit\"at,
Renthof 7, D-35032 Marburg, Germany,
\\E-mail: vacchini@mi.infn.it}


\maketitle\abstracts{
It is argued that
in the description of macroscopic systems inside quantum mechanics the
study of
the dynamics of selected degrees of freedom slowly varying on a
suitable time scale, corresponding to relevant observables
for the given reduced description, is particularly
meaningful. A formalism developing these ideas in the more
simple case of a  microsystem interacting with a
macroscopic system is briefly outlined, together with an
application to the field of neutron optics. The obtained
reduced description relies on a T-matrix formalism and has
the property of complete positivity.
}

\section{Introduction}
More than half a century has passed since Erwin
Schr\"{o}dinger introduced for the first time his by now
celebrated and extensively studied wave equation.
An equation whose interpretation was from the very beginning
problematic. During all these years quantum mechanics has
proven to be strikingly successful and has provided
the explanation for marvelous experiments.
The relativistic extension of this theory, quantum field
theory, has also led to amazing achievements, both with
regard to experimental precision and to the understanding of
nuclear and subnuclear structures. Still,
one cannot feel actually satisfied, due to the fact
that
there are still
great conceptual difficulties in the understanding of the
foundations of quantum mechanics.
And also quantum field theory is burdened with very serious
interpretative difficulties, only partially circumvented by
the useful recipe of renormalization.
Many formal and interpretative schemes have been proposed,
but neither seems to be prevailing or liable to be
definitely proved or disproved by realizable experiments.
It is not even clear
what notions and objects should be taken as fundamental;
a lack of rigor and clarity is felt to undermine the whole
theory, and in particular the problem of
measurement.~\cite{Bell}
Quantum mechanics is said to be the theory of microsystems,
but taking well-known experimental evidences into
account
one is lead to realize that, contrary
to what is often tacitly believed, no direct
objectivity can be attributed to   microsystems, such as for example
particles.
This should be clear if one considers the manifestations
of wave-particle duality; the existence of quantum correlations
which,
as stressed at the very beginning of quantum mechanics by
Schr\"{o}dinger himself,~\cite{Erwin} through the phenomenon
of entanglement ({\it Verschr\"{a}nkung}) make the
attribution of properties to part of a system problematic;
the famous E.P.R. paradox;~\cite{epr} recent experiments in which
the particle picture seems to lead to inconsistencies, e.g., the
heavily debated superluminal photonic tunneling experiments.~\cite{Nimtz}
The dissatisfaction with this situation and the necessity to reconsider
the notion of particle has been
recently stressed also by Haag, who has proposed to take as
fundamental the notion of {\it event}.~\cite{Haag}
A possible alternative approach was elaborated by
Ludwig~\cite{Foundations} (for a brief but
self-consistent survey of Ludwig's axiomatic approach
see also~\cite{phd}): according to his
axiomatic foundations of  quantum mechanics
the basic elements of reality are not
microsystems, but rather the macroscopic setup of any real experiment, which
he divided in preparation and registration apparatuses.
His approach gives a solid basis to the point of view,
initially expressed by Bohr, according to which
the internal coherence of quantum mechanics and
closeness to experimental reality demand that microsystems
should be anchored to the objective reality of macroscopic
systems.
About the connection between the quantum and the classical
description let us only mention a recent review on the
subject,~\cite{Kiefer} paying particular attention to the
problem of decoherence, together with a recently proposed
approach, in which quantum and classical observables are
jointly considered and a notion of {\it event} is also
introduced.~\cite{Jadczyk}

\section{Time Scale and  Macroscopic Systems}
Sharing Ludwig's viewpoint one should start with a phenomenological,
objective, and in this sense classical, description of
macrosystems, macroscopic exactly in the sense that they are
liable to be objectively described.
In particular Ludwig envisaged this objective description in
terms of trajectories for suitable observables or parameters
connected to the system.
Such a description is however still lacking, even though
much progress has been made thanks to the theory of
continuous measurement,~\cite{continue}
mathematically based on the theory
of stochastic processes, which has led to the introduction
of the notion of trajectory in quantum mechanics.
Indeed the very definition of a finite isolated macroscopic
system is slippery, because of the existence of quantum
correlations.
The way in which isolation from the environment is obtained
belongs, in our opinion, to the very definition of the
system.
If one does not take some
approximations into account, the
concept of isolated system can only be an asymptotic one.
Considering
a finite preparation time means that some memory loss
is operatively necessary,
the price of some coarse graining of the dynamical
description must be paid: to do this we associate in a systematic
way to the preparation procedure a suitable
time scale. The relevant role of the preparation procedure means
a breaking of basic space-time symmetry by suitable boundary
conditions which introduce the peculiarities of the system,
hiding the more universal behavior of local or short range
interactions. The field theoretical approach, that is anyway
mandatory in the relativistic case, is best suited to express the
interplay of local universality and peculiar boundary conditions.
The time scale has to be
long enough in order to break up the correlations with the environment
and make the idealized boundary
conditions physically meaningful.
On this time scale one considers the subdynamics of suitable
slow variables. According to the level of description,
the fundamental fields may be associated to
molecules as fundamental constituents or, in a more refined
description, to nuclei and electrons.
The physically relevant observables, slowly varying on the
given time scale, typically densities of conserved charges,
should be connected to the objective properties to be
ascribed to the system.
The time scale associated to the   preparation procedure,
necessary in order to actually define and isolate the
system, accounts for irreversibility, reflected in the
structure of the equations for the relevant variables and
connected to the directedness between  preparation and
registration.
In a completely sharp description of the dynamics of a
subsystem the physics of the whole universe would enter,
correlations could not be neglected.
The proposal is to tune the formalism of  quantum mechanics
to this situation, emphasizing
already in the formalism that only coarse grained descriptions make 
sense: obviously the striving to lower the time scale and to push cutoffs
farther still remains, but should not be based only on formal 
procedures like thermodynamic limit and renormalization.

A significant achievement for the concrete realization of this
research program would be the development of a general
formalism, inside non relativistic quantum field theory, for
the description of the reduced dynamics of slowly varying
degrees of freedom.
Such a description should be meaningful
on a time scale determined by the choice of relevant
observables.
A first elaboration of a formalism with these features
has already been developed in the
case of a  microsystem interacting with a  macroscopic
system~\cite{art1} and will be the
object of the following paragraphs, paying particular attention to
structural properties such as complete positivity.
This formalism will prove suitable for the description of
both coherent and incoherent interactions, as we shall see
in the last paragraph, with reference to the case of neutron
optics. The possibility of describing incoherent effects
being strictly connected to the use of a statistical
operator formalism.
This formal approach has been pursued further in order to
apply it to  macroscopic systems~\cite{be-jp} (see also the
contribution of Prof.~L.~Lanz to these Proceedings).
In this case the reduced
dynamics pertains to some degrees of freedom (e.g.,
distribution function in a kinetic description; densities of
mass, energy and momentum in a hydrodynamic description)
that are slowly varying on the chosen time scale, much
longer than the typical time of microphysical interactions.
The obtained equations are formally very similar to those
derived for the case of the microsystem, so that a kind of
unified description may be envisaged.
This could be a promising feature in connection with the
description of many-body systems in which a coherent
dynamics plays a relevant role, as it happens for the
recently observed Bose-Einstein condensates of trapped
alkali atoms.~\cite{Anderson}
It appears that, considering slow variables, the time
evolution satisfies a generalization of the complete
positivity property.

\section{Subdynamics and Complete Positivity}
\subsection{A Particle Interacting with Matter}
To obtain a  concrete realization of the previously
introduced ideas in a tractable case
we consider the simplest example of subdynamics of a  macrosystem: a
particle interacting with matter
at equilibrium.
This example can be of particular physical interest in
connection with recent so called single particle experiments
using massive particles.
In this case the
subdynamics of the microsystem may be extracted, as a selected degree of
freedom, from the dynamics of the whole system. It can also be seen
as the slightest disturbance to the equilibrium state, a first
tiny step towards the study of non-equilibrium systems.
We will briefly sketch here only the general scheme,
referring the reader to the original paper.~\cite{art1}
Before doing this however, we will
introduce the definition of the property of complete
positivity, so as to fully appreciate its appearance in the
structure of the generator of the dynamical evolution.

\subsection{Complete Positivity}
The most general representation of the preparation
of a physical system
described in a Hilbert space ${\cal H}$ is given by
a statistical operator, that is to say
an operator in the space
${\cal TC}({\cal H})$
of trace class operators on ${\cal H}$, positive and with
trace equal to one.
Consider now a mapping ${\cal U}$ defined on the space of
trace class operators into itself
$
        {\cal U}:
        {\cal TC}({\cal H}) \longrightarrow
        {\cal TC}({\cal H})
$,
possibly corresponding to a
Schr\"odinger picture description on the states.
We say that the map ${\cal U}$ is completely positive, or
equivalently has the property of complete
positivity,~\cite{Kraus-Alicki} if and only if the adjoint map ${\cal U}'$
acting on the space ${\cal B}({\cal H})$ of bounded  linear
operators, dual to ${\cal TC}({\cal H})$,
$
        {\cal U}':
        {\cal B}({\cal H}) \longrightarrow
        {\cal B}({\cal H})
$,
and therefore corresponding to an Heisenberg picture description in terms of
observables, satisfies the inequality
        \begin{equation}
        \label{complete positivity}
        \sum_{i,j=1}^n
        \langle\psi_i\vert
        {\cal U}'
        (
        {\hat{\sf
        B}}{}_i^{\scriptscriptstyle\dagger}
        {\hat {\sf B}}{}_j
        )
        \vert\psi_j\rangle
        \geq 0
        \qquad
        \forall n\in{\bf N}, \quad
        \forall
        \left \{
        \psi_i
        \right \}
        \in {\cal H}
        , \quad
        \forall
        \{
        {\hat {\sf B}}_i
        \}
        \in {\cal B}({\cal H})   .
        \end{equation}
For $n=1$ one recovers the usual notion of positivity, while
for bigger $n$ this is actually a nontrivial requirement.
It is immediately seen that any unitary
evolution is completely positive. In this sense one can see complete positivity as a property
that is worth retaining when shifting from the unitary
dynamics for closed systems to a more general dynamics for
the  description of open systems. In fact the general
physical argument for the introduction of complete positivity is the
following.
Consider a system ${\cal S}_1$ described in ${\cal H}_1$,
whose dynamics is given by the family of mappings
        \[
        {\cal U}:
        {\cal TC}({\cal H}_1) \longrightarrow
        {\cal TC}({\cal H}_1)
        \]
and an $n$-level  system ${\cal S}_2$ described in ${\cal
H}_2 = {\bf C}^n$, whose dynamics can be neglected, so that
${\hat {\sf H}}_2=0$. Because the two  systems do not interact,
the map $\tilde{{\cal U}}$ describing their joint evolution
        \[
        \tilde{{\cal U}}:
        {\cal TC}({\cal H}_1 \otimes {\bf C}^n) \longrightarrow
        {\cal TC}({\cal H}_1 \otimes {\bf C}^n)
        \]
will be simply given by the tensor product
$\tilde{{\cal U}} = {\cal U} \otimes {\bf 1}$. But the
dynamical map  $\tilde{{\cal U}}$ must of course be positive
and this is equivalent to the requirement that ${\cal U}$ be completely positive.
\par
The property of complete positivity has already shown to be particularly relevant in the
determination of quantum structures, for example in the field of quantum
dynamical semigroups, used for the description of
the irreversible dynamics of open quantum
systems, typically the reduced dynamics of systems
interacting with an external system, such as a heat bath or
a measuring instrument.
In Heisenberg picture quantum dynamical semigroups
are given by collections of positive mappings
        \[
        {\cal U}'_t:
        {\cal B}({\cal H}) \longrightarrow
        {\cal B}({\cal H})
        \qquad t\geq 0,
        \qquad
        {\cal U}'_0 {\bf 1}={\bf 1}
        \]
which satisfy the
semigroup composition property
        \[
        {\cal U}'_s
        {\cal U}'_t
        =
        {\cal U}'_{s+t}
        \qquad s,t\geq 0
        .
        \]
Under these conditions a generally unbounded generator ${\cal
L}'$ exists such that
        \[
        {
        d
        \over
         dt
        }
        {\cal U}'_t {\hat {\sf B}}
        =
        {\cal L}'{\cal U}'_t {\hat {\sf B}}
        \]
for all ${\hat {\sf B}}$ in the domain. If one further asks
the semigroup to be norm continuous, so that the generator
is a
bounded map, it can be shown, as has been done by
Lindblad,~\cite{Lindblad} that complete positivity determines the
general expression for the
generator to be of the form
        \[
        {\cal L}' {\hat {\sf B}}
        =
        {i \over \hbar}
        [
        {\hat {\sf H}},
        {\hat {\sf B}}
        ]
        -{1\over 2}
        \biggl\{
        \sum^{}_j
        {\hat {\sf V}}{}_j
        {\hat {\sf V}}{}_j^{\scriptscriptstyle \dagger}
        , {\hat {\sf B}}
        \biggr\}
        +  
        \sum^{}_j
        {\hat {\sf V}}{}_j^{\scriptscriptstyle
        \dagger}
        {\hat {\sf B}}
        {\hat {\sf V}}{}_j
        \]
        \[
        {\hat {\sf V}}{}_j,
        \sum^{}_j
        {\hat {\sf V}}{}_j
        {\hat {\sf V}}{}_j^{\scriptscriptstyle \dagger}
        \in {\cal B}({\cal H}),
        \qquad
        {\hat {\sf H}}{}_j =
        {\hat {\sf H}}{}_j^{\scriptscriptstyle \dagger}
        \in {\cal B}({\cal H})
        \]
This Lindblad structure of master equation, possibly allowing for
unbounded  operators or even quantum fields,
appears in many applications
in very different fields of physics and is often taken as
starting point for phenomenological approaches.
It accounts for a non-Hamiltonian dynamics and has
been extensively used in the
formulation of continuous measurement theory and especially
in quantum optics.

\subsection{Structure of the Generator}
We now come back to the description of the dynamics of a
particle interacting with a macroscopic system, typically
matter at equilibrium, and
consider the total Hamiltonian in the field formalism of
second quantization.
The Hamiltonian contains the
term describing the free
particle $H_0$, the contribution of
matter at equilibrium $H_{\rm m}$ and an interaction
potential $V$
        $$
        {H}={H}_0 + {H}_{\rm m} + {V}
        \qquad  
        \qquad  
        {H}_0 = \sum_f
        {E_f} {a^{\scriptscriptstyle \dagger}_{f}} {a_{{f}}}
         \qquad  
         \qquad  
        {\left[{{a_{{f}}},{a^{\scriptscriptstyle 
        \dagger}_{g}}}\right]}_{\mp}=\delta_{fg} 
        $$
where ${a_{{f}}}$ is the destruction operator for the microsystem,
either a Fermi or a Bose particle, in  
the state $u_f$.
Having it in mind to describe situations in which only one particle
is observed in each experimental run  we assume for the
statistical operator the form
        $$
        {\rho}=
        \sum_{{g} {f}}{}  
        {a^{\scriptscriptstyle \dagger}_{g}}  
        {{\varrho}^{\rm m}} {a_{{f}}}
        {{\sf w}}_{gf}    ,
        $$
where ${{\varrho}^{\rm m}}$ is the  statistical operator
describing matter,
while ${{\sf w}}_{gf}$ is
a matrix with
positive entries and trace one, which can be considered as the
representative of a statistical operator ${{\hat {\sf w}}}$
in the one particle Hilbert spaces ${{{\cal H}^{(1)}}}$.
To understand this choice consider the charge
$Q=\sum_f
        {a^{\scriptscriptstyle \dagger}_{f}} {a_{{f}}}$.
${{\varrho}^{\rm m}}$ is an eigenvector
of this operator with eigenvalue zero, i.e., it contains no
microsystems, while $\rho$ has eigenvalue one, corresponding to a
single  microsystem.
To extract the subdynamics of the microsystem we consider observables
bilinear in the field operators, ${A}
        = \sum_{f,g}  
        {a^{\scriptscriptstyle \dagger}_{f}}  
        {\sf A}_{fg}
{a_{g}} $,
and the following simple reduction formula
        \begin{equation}
        \label{ridotta}
        {\hbox{\rm Tr}}_{{\cal H}}
        \left(  
        {{A}{\rho}}
        \right)  
         = \sum_{f,g} {\sf A}_{fg}
        {{\sf w}}_{gf}=
        {\hbox{\rm Tr}}_{{{\cal H}^{(1)}}}
        \left(
        {{\hat {\sf A}} {\hat {\sf w}}}
        \right)            ,
        \end{equation}
connecting the expectation value of such
observables with $\rho$ to the expectation value in the one particle
Hilbert space of the state and observable corresponding to the given
matrixes.
To develop the calculations one goes over to the Heisenberg
picture
and exploits a superoperator formalism, so that to the T-matrix is
associated the following superoperator ${{\cal T}(z)}$, the
prime denoting superoperators on
$\cal B ({{{\cal H}}})$, the
conjugate space of
$\cal T ({{{\cal H}}})$
        $$
        {\cal T}(z)
        \equiv  
        {\cal V}' + {\cal V}'{{  
        \left(  
        {{ z - {\cal H}'}}  
        \right)  
        }^{-1}}{\cal V}'
        \qquad
        {\cal H}'={i \over \hbar} [{H},\cdot],
        \qquad
        {\cal V}'={i \over \hbar} [{V},\cdot]
        .
        $$
As a result we obtain the following structure for
the evolution mapping on a time $t$ which is small with
respect to the particle's dynamics, though much larger than
the relaxation time of the  macrosystem
        \[
        {{\cal U}'(t)}
        \left(  
        {{{ a}^{\scriptscriptstyle \dagger}_{h}}{{ a}_{k}}}
        \right)  
        =
        {e^{{\cal H}'\tau}}  
        \left(  
        {{a^{\scriptscriptstyle \dagger}_{h}}{a_{k}}}  
        \right)  
        =  
        {{{ a}^{\scriptscriptstyle \dagger}_{h}}{{ a}_{k}}}
        +
        t {\cal L}'
        {{{ a}^{\scriptscriptstyle \dagger}_{h}}{{ a}_{k}}}
        \]
where the generator restricted to this typical bilinear
structure of field  operators in the quasi-diagonal case is
given by:
        \[
        {\cal L}'
        \left(  
        {{{ a}^{\scriptscriptstyle \dagger}_{h}}{{ a}_{k}}}
        \right)  
        =
        {i\over\hbar}  
        \left[  
        { H}_0 + { V},
        { a}^{\scriptscriptstyle \dagger}_{h}
        { a}_{k}
        \right]  
        - {1\over \hbar}  
        \left\{
        \left[  
        { \Gamma} , { a}^{\scriptscriptstyle\dagger}_h
        \right]  
        { a}_k
        -  
        { a}^{\scriptscriptstyle\dagger}_h
        \left[
        { \Gamma}, { a}_k
        \right]  
        \right\}
        +  
        {1\over\hbar} \sum_\lambda  
        { R}_{h \lambda}^{\dagger}
        { R}_{k \lambda}
        \]
${ V}$ and ${ \Gamma}$ being linked respectively to
the
self-adjoint and anti-self-adjoint part of the T-matrix. Let
us note that due to the presence of the minus sign the term
between curly brackets cannot be rewritten as a simple
commutator. Complete positivity of the mapping ${\cal U}' (t)$ restricted to
these simple bilinear field structures
        \[
        \sum_{i,j=1}^n
        \langle\psi_i\vert
        {\cal U}'(t)
        \left(
        \sum_{hk}
        a^{\scriptscriptstyle\dagger}_h
        \langle h\vert
        {\hat{\sf
        B}}{}_i^{\scriptscriptstyle\dagger}
        {\hat {\sf B}}{}_j
        \vert k \rangle    a_k
        \right)
        \vert\psi_j\rangle
        \geq 0
        \]
can be seen from the decomposition which holds true for
an infinitesimal positive time $dt$
        \begin{eqnarray*}
        a^{\scriptscriptstyle\dagger}_h a_k
        \!
        +
        \!
        dt
        {\cal L}'
        (
        a^{\scriptscriptstyle\dagger}_h a_k
        )
        &=&
        \\
        &=&
        {
        \left \{
        {
        a_h+\frac{i}{\hbar}
        dt
        \left[
        { H}_0 + { V}
        ,a_h
        \right]
        -\frac{dt}{\hbar}
        \left[
        { \Gamma} , { a}_h
        \right]  
        }
        \right \}
        }^{\dagger}
        \\
        &&
        \times
        \left\{
        {
        a_k+\frac{i}{\hbar}
        dt
        \left[
        {
        { H}_0 + { V}
        ,a_k
        }
        \right]
        -\frac{dt}{\hbar}
        \left[  
        {
        { \Gamma} , { a}_k
        }
        \right]
        }
        \right\}
        {}+ \frac{dt}{\hbar} \sum_{\lambda}
        { R}_{h \lambda}^{\dagger}
        { R}_{k \lambda}
        \end{eqnarray*}
One can also check that particle number conservations holds,
so that
${\cal L}'({ N})=0$, where ${{ N}}=\sum_f {{a^{\scriptscriptstyle
        \dagger}_{f}}{a_{f}}}$.

Exploiting the
reduction formula  (\ref{ridotta})  we recover a Lindblad equation for the
time evolution of the statistical operator describing the microsystem,
in which the effective Hamiltonian contains a contribution linked to
the self-adjoint part of the T-matrix, averaged over the state of
matter, the gamma  operator being connected instead to its
anti-self-adjoint part
        \begin{equation}
        \label{Lind}
        {
        d 
        \over  
                      dt
        } {\hat {\sf w}}
        =  
        -{i \over \hbar}  
        \left[{\hat {{\sf H}}}_0
        +  
        {\hat {\sf V}},
        {\hat {\sf w}}
        \right]
        -{1\over\hbar}  
        \left \{  
        {  
        {\hat {\sf \Gamma}}
        , {\hat {\sf w}}
        }
        \right \}
        +  
        {1 \over \hbar}  
        \sum^{}_{{\xi\lambda  }}
        {\hat {\sf L}}_{\lambda\xi} {\hat {\sf w}}
        {{\hat {\sf
        L}}{}_{\lambda\xi}^{\scriptscriptstyle \dagger}}\ .
        \end{equation}
The last contribution
is typically incoherent, leading from a pure state to a
mixture and can be introduced only in the formalism of the
statistical operator.
Particle number conservation
implies
$
        {\hat {\sf \Gamma}}
        =
        1/2
        \sum^{}_{{\xi\lambda  }}
        {{\hat {\sf L}}{}_{\lambda\xi}^{\scriptscriptstyle \dagger}}
        {\hat {\sf L}}{}_{\lambda\xi}
$.

\subsection{Neutron Optics as an Application}
In recent years there has been a rapidly growing interest in the  
field of particle optics, especially neutron and atom
optics, due to a spectacular improvement
of the experimental techniques, connected to the introduction of  
the single crystal interferometer in the case of
neutrons.~\cite{rauch-scripta}
Such new achievements
provide very important tests verifying the validity of quantum  
mechanics, especially in that it predicts wavelike behavior
even for single microsystems.  
At the same time a new challenge arises, linked to the accuracy  
required in the description of the interaction between the  
microsystem and the apparatus acting as optical device.
The main interest is devoted to the  coherent wavelike
behavior of particles interacting with homogeneous samples
of matter, as can be justified on the basis of the
similarity between a Schr\"odinger equation with an optical  
potential and the Helmholtz wave equation.
The very existence of such an optical description of the  
interaction is far from trivial and strongly depends  on the  
experimental conditions. While the attention has been mostly devoted to
exploiting the optical analogies, very little has been said on
the borderline between the optical regime, in which  coherent  
effects are predominant and a classical wavelike description  
plays a major role, and an  incoherent regime, where  incoherent  
effects, caused by the interaction between the  microsystem and  
the apparatus and showing typical particle-like features, should not be  
neglected.  
This attitude is exemplified in neutron optics by the use of  
the {\em coherent wave} formalism, instead of a reduced
density matrix description, as usually adopted in quantum  
optics.  
We now want to address the question of how to
consistently describe      
both regimes applying the previously deduced master-equation
(\ref{Lind}), mainly following,~\cite{art2} where the
interested reader can find further details.
The  operators appearing in the generator of the time
evolution are linked to
particle-particle interactions, like the Fermi pseudopotential, and  
to properties of the macroscopic system, like the dynamic structure  
function.
The first part of the generator accounts for the
description of the  coherent interaction in terms of optical  
potential and index of refraction well-known in neutron  
optics, the remaining incoherent part  is related to
the dynamic structure function.

As a first step we want to consider  
the  coherent interaction of neutrons with matter and therefore we  
neglect in (\ref{Lind}) the last contribution, linked to  
incoherent processes. As we will see later this term implies  
indeed a smaller correction in the case of neutron scattering.  
Adopting the Fermi
pseudopotential~\cite{Sears} to describe the neutron nucleus interaction
the T-matrix takes the form
        \[
        {\hat T}=
        {  
        2\pi \hbar^2  
        \over  
        m  
        }  
        b
        {\int d^3 \! {\bf{r}} \,}
        {\psi}^{\scriptscriptstyle\dagger} ({\bf{r}})
        \delta^3 ({\hat {{\sf x}}}-{{\bf{r}}})
        {\psi} ({\bf{r}})
        \]
a local potential parameterized by the coherent scattering
length $b$.
If we consider only pure states
we come to the following stationary Schr\"odinger equation
        \begin{equation}  
        \label{sears}
        \left \{  
        -  
        {  
        \hbar^2  
        \over  
               2m  
        }  
        \Delta_x  
        +  
        {  
        2\pi \hbar^2  
        \over  
        m  
        }     b  
        {\langle  
        {\psi}^{\scriptscriptstyle\dagger}({\bf{x}})
        {\psi}({\bf{x}})
        \rangle}  
        \right \}  
        \phi({\bf{x}})
        = E  
        \phi({\bf{x}})
        ,  
        \end{equation}  
with a potential depending on the average particle density.
If the
medium can be considered homogeneous, with density $n_{\rm {o}}$,
Eq.\ (\ref{sears}) describes propagation of matter waves with an
index of refraction given by
        \begin{equation}
        \label{Gold}
        n \simeq
        [
        1- (\lambda^2 / 2\pi) b n_{\rm{o}}
        ].
        \end{equation}
This leads to the formula currently used to
calculate phase shifts in neutron interferometry  
experiments
        \[
        e^{i \chi}  
        =  
        e^{i (n-1) {2\pi \over \lambda}D}  
        =  
        e^{-i n_{\rm {o}} b \lambda D}  ,
        \]
where $D$ is the thickness of the sample.  

We now come to
the connection between  
the  contributions other than the commutator in  (\ref{Lind})
and the dynamic structure function,  
together with the relevance of this relationship to the optical theorem.
An expression of the form (\ref{Gold})
for the refractive index doesn't include the
contribution to the attenuation of the  coherent wave in the  
medium due to diffuse scattering and hence violates the
optical theorem of scattering theory.
To keep also the attenuation of the coherent wave into
account we have to consider all  contributions in
(\ref{Lind}). Let us  stress from the very beginning some
general features of this expression, thanks to which it can  
describe more general physical situations than those arising  
in an evolution driven by a Schr\"odinger-like equation. The  
last two terms
        \[
        {}-{  
        1  
        \over  
          \hbar  
        }  
        \left \{  
        \frac 12  
        \sum^{}_{{\xi,\lambda  }}  
        {\hat {{\sf L}}}{}_{\lambda\xi}^{\scriptscriptstyle \dagger}  
        {\hat {{\sf L}}}_{\lambda\xi} , {\hat {\sf w}}
        \right \}  
        +  
        {  
        1  
        \over  
         \hbar  
        }  
        \sum^{}_{{\xi,\lambda  }}  
        {\hat {{\sf L}}}_{\lambda\xi}  {\hat {\sf w}}
        {\hat {{\sf L}}{}_{\lambda\xi}^{\scriptscriptstyle \dagger}}  
        \]
allow for the presence of a non-self-adjoint  
potential which is nevertheless not linked to real  
absorption.
This is the case for the present treatment, in
which the imaginary part of the optical potential is to be  
traced back to the existence of diffuse scattering, as  
opposed to the  coherent wavelike behavior. Attenuation of  
the coherent wave is due to the presence of the
anticommutator term,  
responsible for the imaginary potential,  
balanced by the last  contribution,  
typically incoherent in that it leads from a pure state to a  
mixture.  
This last term is given by a sum over subcollections,  
formally similar to the expression that we would obtain for  
the statistical operator after the measurement of a given  
observable. The subcollections are denoted
by the indexes $\lambda\xi$, which specify a change of the  
state of the macroscopic system, caused by interaction with  
the microsystem, thus making this contribution to the  
dynamics incoherent.~\cite{art1}
In fact the trace of this term gives
all the contributions to incoherent scattering, that is to  
say the total diffusion cross section; if the momentum distribution of the
incoming particle is
suitably peaked around ${{\bf{p}}_0}$,
this trace may be written
in the static limit
        \[  
         n_{\rm {o}} b^2 {p_0 \over m}
        \int d\Omega_q  \, S_{\rm {c}} ({\bf{q}})=
        n_{\rm {o}} {p_0 \over m} \sigma_{\rm {d}}
        \]  
where  
        \[  
        S_{\rm {c}} ({\bf{q}})
        =  
        {  
        1  
        \over  
          N  
        }  
        {\int d^3 \! {\bf{x}} \,}
        e^{  
        i{\bf{q}}\cdot{\bf{x}}
        }  
        {\int d^3 \! {\bf{y}} \,}
        \left \langle  
        \delta N({\bf{y}})
        \delta N({\bf{x}}+{\bf{y}})
        \right \rangle         ,  
        \]  
and we have denoted in the structure function
$ S_{\rm {c}} ({\bf{q}}) $
by ${\bf{q}}$ the momentum
transfer and by $\sigma_{\rm {d}}$ the total diffusion cross
section per particle. This is the result derived~\cite{Sears} for the
attenuation of the  coherent beam due to incoherent  
scattering, usually obtained by an evaluation of the local
field effects, neglected in the equation giving the optical neutron  
dynamics (\ref{sears}). In this
approach the incoherent
contribution is already present in the equation giving the  
dynamics of the  microsystem, thanks to the more general
formalism adopted.
The correction to the optical potential
can be read by
        \[
        {\hat {{\sf U}}}  =  
        {  
        2\pi \hbar^2  
        \over  
                    m  
        }  
        n_{\rm {o}}
        \left[  
        b -i  
        {  
        b^2  
        \over  
           4\pi  
        }  
        {p_0\over \hbar}  
        \int d\Omega_q \,  
        S_{\rm {c}}({\bf{q}})
        \right]  
        \]
and is of second order in the
small parameter $b$.

The  incoherent  contribution is thus
necessary to fulfill the optical theorem and take diffuse  
scattering, that attenuates the  coherent beam, into account.
Even though it introduces a smaller correction the  incoherent
contribution is very important from the theoretical point of  
view.
It is not surprising that
the incoherent contribution to the dynamics
has grown out of a thoroughly quantum mechanical  
treatment, as shown by the typical quantum structure of the  
Lindblad equation, relying on non-commutating   
operators, in which an essential role is played by the  
statistical operator ${\hat {\sf w}}$, rather then by
the wave function $\psi$. This point is of central relevance,  
since the  terms which describe the  incoherent  
dynamics cannot be introduced in the  formalism of the wave  
function and are therefore unavoidably absent in an optical-like  
treatment, simply reminiscent of classical optical descriptions.  

\section*{Acknowledgments}
I am very indebted to Prof.~L.~Lanz for careful reading of
the manuscript and precious suggestions. I would
like to thank Prof.~O.~Melsheimer
for helpful discussions and also for kind
hospitality at the University of Marburg.
This work has been supported by the Alexander von
Humboldt-Stiftung.

\end{document}